\documentclass[prx,twocolumn,preprintnumbers,superscriptaddress,longbibliography, 10pt]{revtex4-2} 

\usepackage{graphicx}
\usepackage{bm}
\usepackage{amsmath}
\usepackage{amssymb}
\usepackage{amsfonts}
\usepackage{slashed}
\usepackage{latexsym,epsfig,bbm}
\usepackage{algorithm}
\usepackage{algorithmic}

\newcommand{\bra}[1]{\langle{#1}|}
\newcommand{\ket}[1]{|{#1}\rangle}

\newcommand{\braket}[2]{\langle{#1}|{#2}\rangle}
\newcommand{\bopk}[3]{\langle{#1}|{#2}|{#3}\rangle}

\newcommand{\figref}[1]{Fig.~\ref{#1}}
\newcommand{\figsref}[1]{Figs.~\ref{#1}}

\usepackage{color}
\definecolor{blue}{rgb}{0,0.2,1}

\definecolor{red}{rgb}{0.9,0,0}

\begin{document}

\title{Non-destructively probing the thermodynamics of quantum systems with qumodes}

\author{Thomas J.~Elliott}
\email{physics@tjelliott.net}
\affiliation{Department of Physics \& Astronomy, University of Manchester, Manchester M13 9PL, United Kingdom}
\affiliation{Department of Mathematics, University of Manchester, Manchester M13 9PL, United Kingdom}
\affiliation{Department of Mathematics, Imperial College London, London SW7 2AZ, United Kingdom}
\author{Mile Gu}
\email{mgu@quantumcomplexity.org}
\affiliation{Nanyang Quantum Hub, School of Physical and Mathematical Sciences, Nanyang Technological University, Singapore 637371}
\affiliation{Centre for Quantum Technologies, National University of Singapore, 3 Science Drive 2, Singapore 117543}
\affiliation{MajuLab, CNRS-UNS-NUS-NTU International Joint Research Unit, UMI 3654, Singapore 117543, Singapore}
\author{Jayne Thompson}
\email{thompson.jayne2@gmail.com}
\affiliation{Horizon Quantum Computing, 05-22 Alice@Mediapolis, 29 Media Circle, Singapore 138565}
\affiliation{Centre for Quantum Technologies, National University of Singapore, 3 Science Drive 2, Singapore 117543}
\author{Nana Liu}
\email{nana.liu@quantumlah.org}
\affiliation{Institute of Natural Sciences, Shanghai Jiao Tong University, Shanghai 200240, China} 
\affiliation{Ministry of Education Key Laboratory in Scientific and Engineering Computing, Shanghai Jiao Tong University, Shanghai 200240, China}
\affiliation{University of Michigan-Shanghai Jiao Tong University Joint Institute, Shanghai 200240, China}

\date{\today}

\begin{abstract}
Quantum systems are by their very nature fragile. The fundamental backaction on a state due to quantum measurement notwithstanding, there is also in practice often a destruction of the system itself due to the means of measurement. This becomes acutely problematic when we wish to make measurements of the same system at multiple times, or generate a large quantity of measurement statistics. One approach to circumventing this is the use of ancillary probes that couple to the system under investigation, and through their interaction, enable properties of the primary system to be imprinted onto and inferred from the ancillae. Here we highlight means by which continuous variable quantum modes (qumodes) can be employed to probe the thermodynamics of quantum systems in and out of equilibrium, including thermometry, reconstruction of the partition function, and reversible and irreversible work. We illustrate application of our results with the example of a spin-1/2 system in a transverse field.
\end{abstract}
\maketitle 

\section{Introduction}

Quantum systems are now routinely created, manipulated, controlled and studied in the lab, making the realisation of technologies that exploit truly quantum phenomena a very imminent reality. Examples of such systems are numerous, including ultracold atoms~\cite{bloch2012}, ion traps~\cite{blatt2012}, superconducting circuits~\cite{clarke2008}, and microwave cavities~\cite{raimond2001}. Beyond the intrinsic interest in studying such systems, they can be deployed to, for example, enhance computation and metrology~\cite{monroe2002, nielsen2010, giovannetti2011, haroche2013}. A particularly promising use, especially in the near term, is that of analogue quantum simulation, where the system is made to emulate the behaviour of another quantum system of interest~\cite{lewenstein2012, johnson2014what,daley2022practical}, which can be applied in quantum chemistry~\cite{cao2019quantum} to study otherwise experimentally inaccessible quantum systems and parameter regimes.

A necessary element in these applications is to read-out properties of or results output by the quantum system. While theoretical models of measurement are well-established, in many of the proposed architectures the measurement process is destructive to the system, allowing only a single-shot reading, after which the system must be completely reprepared. For example, with cold atom systems, a widely-used measurement method is time-of-flight~\cite{hadzibabic2003, altman2004, folling2005, bloch2008}, which involves destruction of the atomic trapping potential, hence requiring the atoms to then be re-trapped and re-cooled. Moreover, in other systems direct methods to measure particular sets of observables may not be available.

It has been proposed that this can be mitigated through the use of ancillary systems that serve as measurement probes. Through their interaction, properties of the system are imprinted onto the probes~\cite{recati2005, bruderer2006}, and information about the system may then be obtained through measurement of the ancillary probes alone~\cite{goold2011, knap2012, haikka2013, mazzola2013, dorner2013, sabin2014, fusco2014, batalhao2014, hangleiter2015, cosco2015, elliott2016, johnson2016, bermudez2017, benedetti2017}. While such measurements disturb the system state and are hence not non-demolition, they do not destroy the system. They thus constitute a practical method to determine certain system properties whilst leaving the system intact. This method has been explored in particular for cold atom systems, where atomic impurities form ancilla qubits~\cite{bentine2017}; post-processing of the impurity measurement statistics then allows properties such as density~\cite{elliott2016} and temperature~\cite{sabin2014, johnson2016} to be determined.

In parallel, developments in continuous variable quantum information processing~\cite{lloyd1999, braunstein2005, furusawa2011quantum, weedbrook2012, andersen2015}, based on continuous variable quantum modes (`qumodes') rather than qubits, provide new applications for quantum optics and collective atomic phenomena in quantum technologies. One recently proposed model of quantum computation uses squeezed qumodes as a resource for phase estimation of an operator~\cite{liu2016}.

Bringing these two themes together, qumodes offer a promising, flexible means of non-destructively probing quantum systems. With an appropriate initial qumode state, the statistics of the system operator to which the qumode couples can be mapped directly onto the qumode state. Subsequent measurement of the qumode then allows for the spectrum of the system operator to be determined, along with the populations of the respective eigenstates. This enables a full characterisation of the moments of the operator for the system state, as though one had directly sampled the observable from the system. In the ideal case, the qumode behaves akin to a non-destructive von Neumann measurement meter~\cite{neumann1932mathematische, brune1996}, while even with a more realistic model of achievable initial qumode states a highly faithful measurement can be performed. Crucially, we only directly measure the qumode ancilla, from which we are able to indirectly probe the system of interest itself, circumventing the typical direct approaches that are often destructive to the system.

Here, we highlight how we can employ these probes to study the thermodynamics of quantum systems. We first describe the basic qumode probing protocol, and how the operator statistics are imprinted onto the qumode state. We show that the protocol is robust to experimental limitations of finite squeezing in the initial qumode state, and discuss its experimental practicality. We then introduce several thermodynamical applications of the probing protocol. Amongst these, we show how qumodes can be used as thermometers to measure the temperature of quantum systems in equilibrium, and how partition functions can be reconstructed. We also show how reversible and irreversible contributions to work can be measured in non-equilibrium settings, extending beyond prior work that introduced an analogous approach to measuring total work~\cite{cerisola2017, ahmad2022finite}. We also show how the protocol can be used to measure the overlaps of ground states of different parameter regimes of a Hamiltonian. As an example, we simulate an illustrative use of our method in probing the temperature of a spin-1/2 system in a transverse field.

\section{Qumode Probes}

\subsection{Base qumode probing protocol}

We first describe the basic qumode probing protocol, from an architecture-agnostic perspective. Consider a quantum system of interest with (possibly unknown) state $\rho_{\mathrm{sys}}$, and an observable $\mathcal{O}$ of the system that we wish to measure. Specifically, we wish to determine the moments of $\mathcal{O}$ with respect to $\rho_{\mathrm{sys}}$, given by $\langle O^m\rangle=\mathrm{Tr}(\rho_{\mathrm{sys}}O^m)$, where $O$ is the operator associated with $\mathcal{O}$. The aim is to make these measurements in a manner non-destructive to the system. This can be achieved by a probing protocol that uses ancillary qumodes as probes. We emphasise that the protocol is not a full tomography of the state or operator~\cite{vogel1989, hradil1997, dariano2003, paris2004quantum}, but rather, a means to obtain the statistics of the observable with respect to the system state.

\begin{figure}
\includegraphics[width=\linewidth]{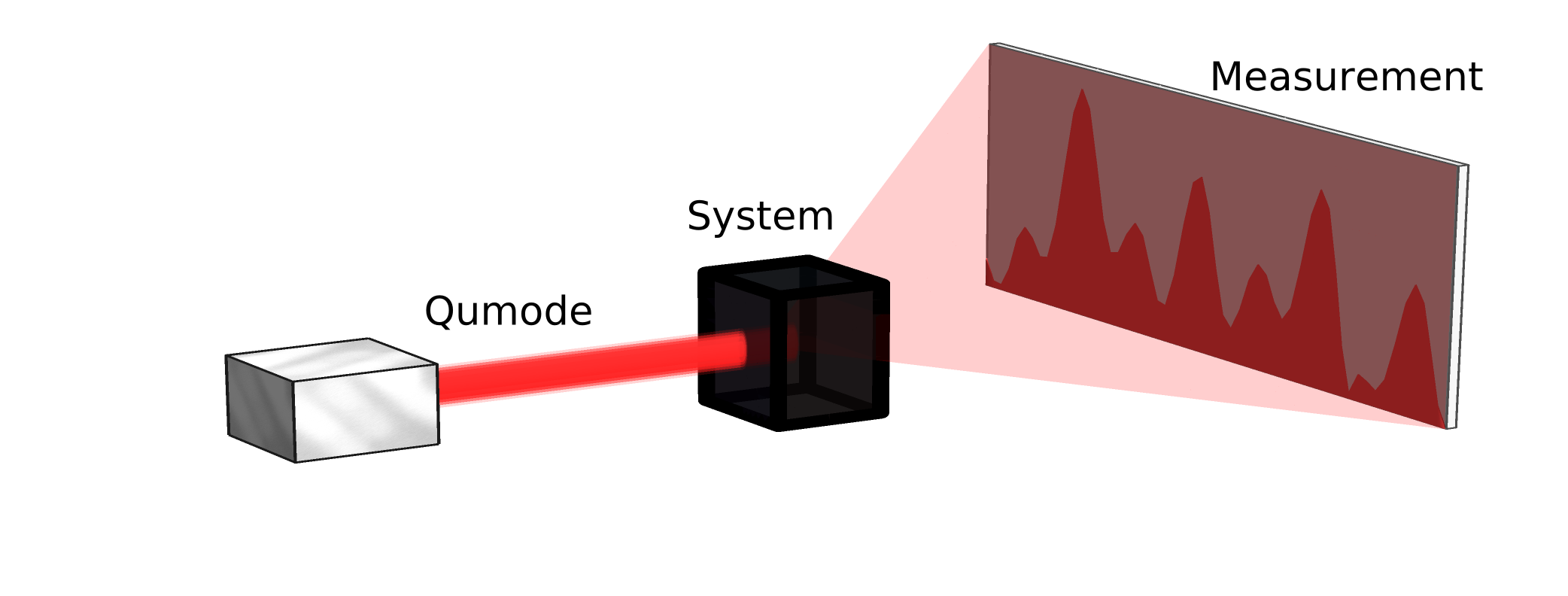}
\caption{{\bf Probing quantum systems with qumodes.} A continuous variable qumode illuminates a system of interest, inheriting properties of the system through their interaction. The state of the qumode is then measured, revealing information about the system.}
\label{figprobe}
\end{figure}

The qumode probing protocol consists of three components (see \figref{figprobe}). The first of these is the system of interest, with state $\rho_{\mathrm{sys}}$. The protocol does not in general need a particular form for the system or its state, and it may inhabit a discrete or continuous Hilbert space. As mentioned above, there is no need for \emph{a priori} knowledge of this state, and in general this nescience is assumed. The second component is the continuous variable qumode, described by its quadratures $x$ and $p$, often referred to as `position' and `momentum'~\cite{gerry2005}. We shall take these quadratures to be in their dimensionless form; that is, in terms of the creation and annihilation operators $a$ and $a^\dagger$ of the mode, we have $x=(a+a^\dagger)/2$ and $p=(a-a^\dagger)/2i$.

The final component is the interaction between system and qumode. We shall consider an interaction Hamiltonian of the form $\lambda x\otimes H_{\mathrm{int}}$, where the first subspace belongs to the qumode, and the second the system~\cite{liu2016}. Hence, the interaction acts on the system, with a strength that depends on the qumode position quadrature, with $\lambda$ an overall coupling strength~\footnote{In principle the coupling strength can have a time dependence. Here we do not consider this beyond the straightforward binary on/off nature of a constant interaction strength.}. The associated evolution operator (in natural units $\hbar=1$) is $U(t)=\exp(-i\lambda x\otimes H_{\mathrm{int}}t)$, and thus the qumode is dephased in this quadrature, at a rate dependent on the the system operator $H_{\mathrm{int}}$, thence motivating the use of a phase estimation-type algorithm.

We label the eigenstates of the system operator $H_{\mathrm{int}}$ as $\ket{u_n}$, with associated eigenvalues $E_n$. Thus, when the system is in such an eigenstate, and the qumode in a quadrature eigenstate $\ket{x}$, the effect of the interaction can be written
\begin{equation}
\label{eqeigenstates}
\ket{x}\otimes\ket{u_n}\to e^{-i\lambda xE_nt}\ket{x}\otimes\ket{u_n}.
\end{equation}

In general, the qumode will be in a superposition of the quadrature eigenstates $\ket{\psi_q}=\int dx G(x)\ket{x}$, and the system state can always be expressed in the basis defined by the eigenstates of $H_{\mathrm{int}}$: $\rho_{\mathrm{sys}}=\sum_{mn}c_{mn}\ket{u_m}\bra{u_n}$. Owing to the linearity of quantum mechanics, Eq.~\eqref{eqeigenstates} can be extended to such states, and one can perform a partial trace over the system to obtain an expression for the qumode state after running the interaction for a time $t$:
\begin{equation}
\label{eqintstate}
\rho_q(t)=\iint dxdx'G(x)G^*(x')L(x,x',t)\ket{x}\bra{x'},
\end{equation}
where analogous to the qubit probing protocols~\cite{elliott2016, johnson2016}, we define the dephasing function 
\begin{align}
L(x,x',t):=&\mathrm{Tr}\left(\rho_{\mathrm{sys}}e^{-i\lambda(x-x')H_{\mathrm{int}}t}\right)\nonumber \\
=&\sum_n P_ne^{-i\lambda(x-x')E_nt},
\end{align}
where $P_n=c_{nn}$. This is resemblant of a characteristic function for the system operator, and is the crux of the probing protocol.

\begin{figure}
\includegraphics[width=\linewidth]{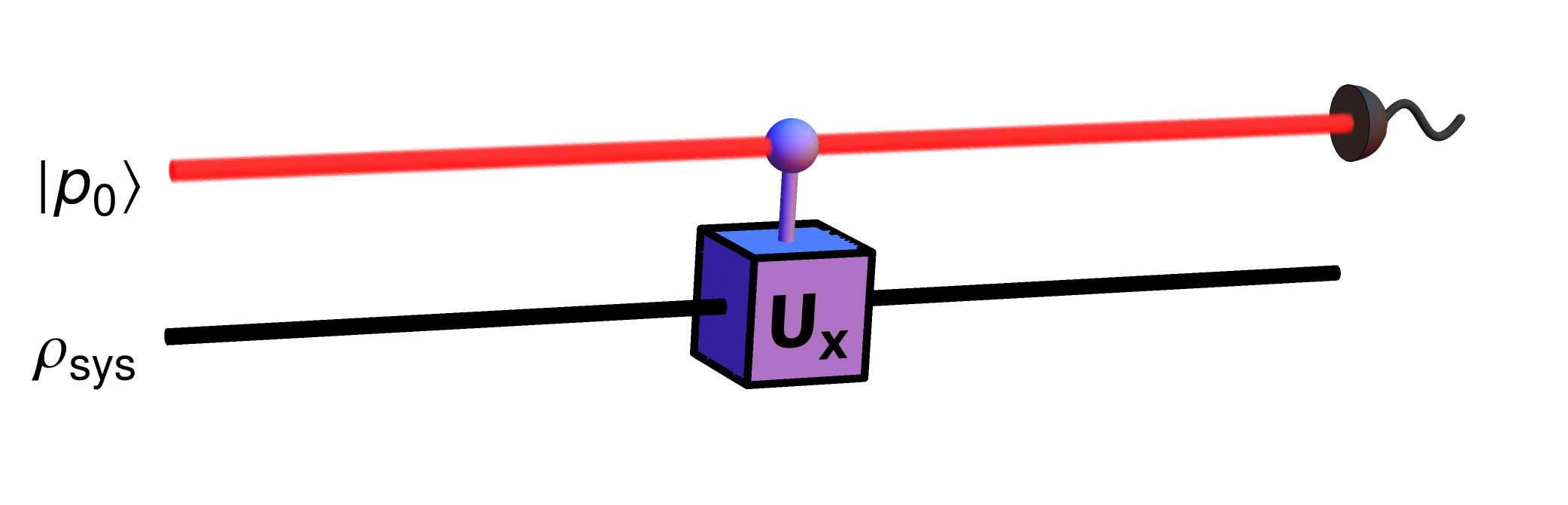}
\caption{{\bf Quantum circuit for qumode probing.} A qumode prepared in momentum eigenstate $\ket{p_0}$ interacts with the system through a controlled gate $U_x=\exp(-i\lambda xH_{\mathrm{int}}t)$ dependent on the qumode position quadrature $x$. Measuring the qumode in the momentum quadrature then directly samples the statistics of the system operator $H_{\mathrm{int}}$ for state $\rho_{\mathrm{sys}}$.}
\label{figcircuit}
\end{figure}

Let us now consider that after an interaction time $\tau$ a measurement is made of the qumode state. Inspired by the qubit-based protocols, we shall measure in a basis conjugate to that which defines the interaction Hamiltonian, here the momentum quadrature. Recall that these can be expressed in terms of the position eigenstates as $\ket{p}=(1/\sqrt{2\pi})\int dx \exp(ipx)\ket{x}$. Correspondingly, using that $P(p):=\bopk{p}{\rho_q(\tau)}{p}$, we have
\begin{align}
\label{eqgenmeas}
P(p)&=\frac{1}{2\pi}\iint dx dx'G(x)G^*(x')e^{-ipx}L(x,x',\tau)e^{ipx'}\nonumber\\
&=\!\!\sum_n\!\frac{P_n}{2\pi}\!\!\iint\! dx dx'G(x)G^*(x')e^{-i(p+\lambda E_n\tau)x}e^{i(p+\lambda E_n\tau)x'}\nonumber\\
&=\sum_nP_n|\mathcal{G}(p+\lambda E_n\tau)|^2,
\end{align}
where $\mathcal{G}(p):=(1/\sqrt{2\pi})\int dx G(x)\exp(-ipx)$ is the Fourier transform of $G(x)$. Thus, we see that the eigenvalues of the interaction Hamiltonian -- and their associated probabilities -- are imprinted into the final state of the qumode.

Before discussing the choice of initial probe state distribution $G(x)$, let us address a caveat to the above derivation. Namely, that we have neglected the presence of the natural evolution of the system under its bare Hamiltonian $H_0$ during the running of the protocol. For this to be valid, we require that the interaction occurs on timescales much faster than the natural evolution ($\lambda H_\mathrm{{int}}\gg H_0$), and that the natural evolution has negligible effect on the system state during the running of the protocol ($H_0\tau\ll1$), hence imposing a maximum allowable running time for the protocol. However, these restrictions are lifted when the bare Hamiltonian and the interaction Hamiltonian commute, in which case the natural evolution does not affect the outcome of the qumode measurement.

\subsection{Initial probe state}

Consider the idealised scenario where the qumode is initially prepared in an eigenstate of momentum $\ket{p_0}$; $G(x)=(1/\sqrt{2\pi})\exp(ip_0x)$. In this case, we find that $\mathcal{G}(p)=\delta(p-p_0)$, leading to 
\begin{equation}
\label{eqinfinitesq}
P(p)=\sum_n P_n |\delta(p-(p_0-\lambda E_n\tau))|^2.
\end{equation}
That is, the final state of the probe has a distribution that consists of sharp peaks, non-zero only at the points $p=p_0-\lambda E_n\tau$, where it takes values $P_n$. The measurement of the qumode state hence directly samples the same distribution as that of a measurement of the operator $H_{\mathrm{int}}$ on the state $\rho_{\mathrm{sys}}$, with the mapping from qumode measurement outcomes to the spectrum of the system operator given by $E=(p_0-p)/\lambda \tau$. With repeated measurements, one can then obtain an estimate of the probability distribution $P(p)$ (and hence $P(E)$). This thus allows for the estimation of moments of the system operator $\langle H_{\mathrm{int}}^m\rangle=\sum_n P_nE_n^m$. Hence, by encoding a suitable operator $O$ associated with some observable of interest $\mathcal{O}$ as our interaction Hamiltonian, we can use the protocol to determine the moments of $\mathcal{O}$ with respect to the system state $\rho_{\mathrm{sys}}$ -- effectively realising an ideal von Neumann measurement meter. It is worth noting that this does not require prior knowledge of the eigenvalues of the system operator, and further, that these can be determined from the qumode measurement outcomes. In contrast to the analogous qubit probing protocols, here all these properties can be obtained directly, without the need for post-processing of the measurement outcomes. This protocol is illustrated in \figref{figcircuit}.

However, this is an idealised model of the protocol, and in practice the qumode cannot be prepared in a true momentum quadrature eigenstate. Rather, one can only achieve approximations to this, with a finite level of squeezing in a given quadrature (the eigenstates corresponding to the limit of infinite squeezing). That is, we have a Gaussian uncertainty in the value of the momentum quadrature, centred on the desired $p_0$, i.e.,
\begin{equation}
G(x)=\left(\frac{s^2}{\pi}\right)^{\frac{1}{4}}\int dq e^{-\frac{s^2q^2}{2}}e^{iqx}\frac{e^{ip_0x}}{\sqrt{2\pi}}.
\end{equation}
Here, $s$ corresponds to the dimensionless squeezing factor~\cite{liu2016}, parameterising the squeezing in the momentum quadrature \footnote{Note that $s=1$ corresponds to the case of an unsqueezed coherent state, defined as the eigenstate of the annihilation operator; $a\ket{\alpha}=\alpha\ket{\alpha}$.}. Inserting this in to Eq.~\eqref{eqgenmeas}, we find the final probability distribution for the qumode is given by
\begin{equation}
\label{eqfinalsqueeze}
P(p)=\frac{s}{\sqrt{\pi}}\sum_nP_n e^{-s^2(p-p_0+\lambda\tau E_n)^2}.
\end{equation}

Indeed, the form of this distribution should not be too surprising -- the smearing of the initial momentum distribution manifests in a smearing in the final distribution. This smearing is identical in the initial and final distributions, and is controlled by the squeezing factor. Specifically, the standard deviation of the final momentum distribution is given by $\varsigma_p=\sqrt{2}/s$, corresponding to a standard deviation for the system operator eigenvalues of $\varsigma_E=\sqrt{2}/s\lambda\tau$. Thus, the precision to which we can measure the eigenvalues of $H_{\mathrm{int}}$ can be increased by running the protocol for longer or increasing the coupling strength, or increasing the squeezing. This is analogous to a result found in the proposal for the power of one qumode computation~\cite{liu2016}, that one can trade off a decreased squeezing with increased running time $\tau$, and vice versa.

While it is tempting to then conclude that these uncertainties in the final distribution can thus be negated by a sufficiently increased running time, this is not necessarily practical in general, due to the constraint imposed on $\tau$ for the effects of the system's natural evolution $H_0$ to be neglected -- as well as the inherent difficulties with maintaining coherence over long timescales. Ultimately, such practical considerations bound how small the spread $\varsigma_E$ can be made. 

More generally, suppose we express the initial qumode state in the momentum eigenbasis as $\ket{\psi_q}=\int dpf(p)\ket{p}$. We note that $f(p)$ and $g(x)$ are related by a basis change corresponding to a Fourier transform, i.e., $f(p)=\mathcal{G}(p)$. Thus, we can equivalently express Eq.~\eqref{eqgenmeas} as \mbox{$P(p)=\sum_nP_n|f(p+\lambda E_n\tau)|^2$}; the final distribution is a mixture of shifted versions of the initial momentum distribution, with weights given by the probability of each eigenstate of $H_{\mathrm{int}}$, and shifts proportional to the corresponding eigenvalue.

\subsection{Candidate experimental platforms}

The general form of the coupling Hamiltonian considered, $\lambda x\otimes H_{\mathrm{int}}$, appears almost ubiquitiously in continuous-variable systems, as these typically couple through one of the quadratures of the qumode(s). Perhaps most famously, the basic building block of quantum light-matter interactions is the quantum Rabi model~\cite{gerry2005}:
\begin{equation}
\label{eqquantumrabi}
H_{QR}=\lambda x\otimes\sigma_x,
\end{equation}
where $\sigma_x$ is the usual Pauli $x$ matrix~\cite{nielsen2010}, taking the role of the interaction Hamiltonian $H_{\mathrm{int}}$. The Hamiltonian was originally conceived as a description of a quantised light field interacting with a two-level system. One often finds this Hamiltonian in its simplified guise as the Jaynes-Cummings Hamiltonian, where the approximation is made to neglect the counter-rotating terms $a\sigma^-$ and $a^\dagger\sigma^+$; nevertheless, systems described by this Hamiltonian are typically more accurately described by the full Rabi Hamiltonian.  Moreover, though the Hamiltonian Eq.~\eqref{eqquantumrabi} contains only a $\sigma_x$ coupling, it is possible to probe the spin operator in any chosen direction by an appropriate rotation of the individual spins prior to running the probing protocol, due to the rotation mapping the statistics of the desired direction onto the $x$-axis (e.g.~applying a Hadamard gate~\cite{nielsen2010} to the system allows probing of $\sigma_z$).

Both cavity~\cite{gleyzes2007, hamsen2017} and circuit~\cite{lahaye2009, forn2010, oconnell2010, yoshihara2017} quantum electrodynamics experiments consist of interactions between a continuous variable mode (cavity fields in the former, nanomechanical resonators in the latter) and a two-level system (atoms and superconducting qubits respectively), interacting through a quantum Rabi Hamiltonian Eq.~\eqref{eqquantumrabi} in the (ultra)strong coupling regime. Such setups operate in a regime where the qumode measurement resolution can be much finer than the differences between the interaction Hamiltonian eigenvalues, which for this example is of order unity. For example, in Ref.~\cite{forn2010} the coupling between qubit and resonator gives $\lambda \approx 10^{10}$, and the Q-factor of $10^3$ and resonance frequency of 8.2GHz leads to $\lambda \tau\sim 200$ when the protocol is run for times of the order of the resonator lifetime. With the parameters of Ref.~\cite{hamsen2017}, we would have $\lambda\tau=40$ when $\tau$ is the cavity lifetime. Comparing with our example later, we see that this would offer a high degree of precision, with a very fine resolution of the spectrum. 

While Eq.~\eqref{eqquantumrabi} makes it clear that the protocol can be used to probe moments of the spin operator of a two-level system, it can be applied more generally. First, the physical motivation and derivation of the Hamiltonian does not necessarily require that the system has only two states, and can be rederived for any number of states, by replacing $\sigma_x$ with the appropriate spin operator for the number of states. Secondly, by illuminating an array of such systems with the same light field, the Hamiltonian becomes an interaction between the light field and the sum of the individual spin operators for each system, and thus probes moments of the total spin operator, as well as correlations between individual spins. This can be enhanced by tuning the optical geometry~\cite{elliott2015} such that different spins couple to the qumode with different strengths and phases. Further, when the individual spins are indistinguishable, they couple to the qumode as a single, collective spin operator, leading to the Dicke model $H_D=\lambda x\otimes J_x$~\cite{baumann2010,landig2016}, which may be probed in a similar manner. Finally, higher-order light-matter interactions in the context of many-body systems often appear in such a form~\footnote{Note that this is not always apparent from a cursory inspection of the relevant Hamiltonians, as these are often derived with counter-rotating terms dropped -- though a more full derivation will yield the appropriate form.}, such as in the case of cold atoms in optical cavities~\cite{ritsch2013, elliott2016engineering}.

We emphasise that the `momentum' measurement made of the qumode is that of its canonical momentum quadrature, which does not necessarily correspond to a Newtonian momentum. Measurement of such quadratures is a highly routine procedure for photons in quantum optics~\cite{slusher1985observation}, and can similarly performed for phonon qumodes formed by e.g., atomic or ionic motion~\cite{bastin2006measure, cerisola2017}.

\section{Qumodes as Thermodynamical Probes}

\subsection{States in equilibrium}

Consider a system that is known to be in a state of thermal equilibrium $\rho_\Theta(\beta)=\exp(-\beta H_\Theta)/Z(\beta)$ with respect to a Hamiltonian $H_\Theta$. Here, $\beta=1/T$ is the inverse temperature (we employ units in which Boltzmann's constant $k_B=1$), and $Z(\beta)=\mathrm{Tr}(\exp(-\beta H_\Theta))$ is the partition function. When the system component of the qumode interaction Hamiltonian $H_{\mathrm{int}}$ is proportional to $H_\Theta$~\footnote{Note that in the circumstance where $H_\Theta$ also coincides with the natural Hamiltonian of the system's evolution $H_0$, this also lifts the requirement that the qumode-system interaction takes place on short timescales relative to the natural evolution.}, we can use the qumode as a thermodynamical probe of the state.

Specifically, notice that the qumode probing protocol in this case will reveal for each eigenvalue $\bar{E}_n$ of $H_\Theta$ its respective probability $P_n$ in the state $\rho_\Theta(\beta)$. Given that this eigenvalue has degeneracy $g_n$, we can for each eigenvalue construct an equation of the form
\begin{equation}
\label{eqZeq}
\ln (Z(\beta))+\beta \bar{E}_n=\ln \left(\frac{g_n}{P_n}\right),
\end{equation}
which follows from taking the logarithm of the spectrum of $\rho_\Theta(\beta)$. Let us assume that we know two of the eigenvalues $\bar{E}_{n_0}$ and $\bar{E}_{n_1}$, and their associated degeneracies $g_{n_0}$ and $g_{n_1}$ respectively. By combining the associated Eq.~\eqref{eqZeq} for both these eigenvalues, we obtain $\beta=\ln(P_{n_0} g_{n_1}/(P_{n_1} g_{n_0}))/(\bar{E}_{n_1}-\bar{E}_{n_0})$, i.e.,
\begin{equation}
\label{eqtemp}
T=\frac{\bar{E}_{n_1}-\bar{E}_{n_0}}{\ln\left(\dfrac{P_{n_0} g_{n_1}}{P_{n_1} g_{n_0}}\right)}.
\end{equation}
By using the qumode to determine the associated $P_{n_0}$ and $P_{n_1}$ it thus provides a non-destructive means of determining the temperature of the system. Note that we do not need \emph{a priori} knowledge of $\bar{E}_{n_0}$ and $\bar{E}_{n_1}$, provided that we know the proportionality factor between $H_{\mathrm{int}}$ and $H_\Theta$ -- the spectra $\{E_n\}$ will be similarly proportional to $\{\bar{E}_{n}\}$, and the former can be deduced from the output distribution. We also remark that a known constant shift in the (proportionally scaled) eigenvalues of one Hamiltonian relative to the other can similarly be accounted for.

Unlike traditional means of thermometry~\cite{correa2015individual}, we do not require thermal equilibrium to be established between the system and the probe -- the temperature is measured through the shifts in the qumode state. Further, while similar schemes for measuring temperature using impurity probe qubits~\cite{johnson2016, sabin2014} also do not require equilibriation, their accuracy is limited by the number of probes. This can be a limitation even for quantum thermometers that exploit advantages in precision using quantum metrology~\cite{stace2010quantum, sabin2014}. In contrast, the precision of qumodes can also be enhanced by either increasing the squeezing factor, or increasing the interaction time or strength.

This approach requires that we can resolve between different eigenvalues of the interaction Hamiltonian $\{E_n\}$, i.e., min($|E_n-E_{n'}|)\gtrsim\varsigma_E$), or at least for the two eigenvalues used in Eq.~\eqref{eqtemp} and their nearest neighbours. Otherwise, we must combine nearby eigenvalues and treat them as degenerate, in term limiting the precision to which the temperature $T$ can be estimated by the (scaled) uncertainty $\varsigma_{\bar{E}}$. The number of measurements required to determine the temperature to a given precision will scale inversely proportional to both the variance of the measurements of $E_n$ (set by $s$, $\lambda$, and $\tau$) due to the finite width of the output distribution peaks, and the associated eigenstate probabilities $P_{n_0}$ and $P_{n_1}$ involved in Eq.~\eqref{eqtemp}.

Conversely, suppose we have knowledge of the temperature of the system (either beforehand, or using the above approach), and at least one of the degeneracies $g_{n_0}$. Then, we can use qumode probing to deduce the remaining degeneracies by considering the ratio of the associated probabilities of the equilibrium state. Specifically, we have that
\begin{equation}
g_n=\frac{P_n}{P_{n_0}}g_{n_0}e^{\beta(\bar{E}_n-\bar{E}_{n_0})}.
\end{equation}
From this we are then able to reconstruct the full partition function $Z(\beta)=\sum_n g_n \exp(-\beta \bar{E}_n)$. This offers access to several important thermodynamical quantities of interest~\cite{glazer2002}, either directly, or by approximating derivatives of $Z(\beta)$ if we are able to slightly perturb the temperature. Amongst these quantities are the free energy ($F(\beta)=-\log(Z(\beta))/\beta$), heat capacity ($C=\beta^2 \partial^2 \log(Z(\beta))/\partial \beta^2$), and entropies ($S_\alpha=\log\mathrm{Tr}(\rho_\Theta(\beta)^\alpha)/(1-\alpha)$) -- the latter two of which can be used to probe quantum critical points~\cite{fradkin2006entanglement,liang2015}. 

Particularly valuable is the access reconstruction of $Z(\beta)$ provides to the free energy landscape of the system. Previously, proposals have been introduced, based either on the two-time measurement method, or using a qudit-~\cite{paz2014} or qumode-based~\cite{cerisola2017, ahmad2022finite} method, to sample from the distribution of work done on a system $P(W)$ between two timepoints, where $W$ corresponds to the difference in energy before and after the evolution (indeed, the latter of these can be seen as an important special case application of our qumode-probing framework). Then, using the quantum counterpart~\cite{tasaki2000jarzynski, mukamel2003quantum} of the Jaryzinski equality~\cite{jarzynski1997nonequilibrium}, the free energy difference is deduced from the work done -- $\langle\exp(-\beta W)\rangle=\exp(-\beta \Delta F)$. However, these methods are not always efficient, particularly at low temperatures, or when large negative values of work are involved~\cite{paz2014}.In contrast, our method based on probing $\{E_n\}$ and $\{P_n\}$ directly still allows for the free energy to be recovered efficiently in those regimes.

\subsection{Non-equilibrium thermodynamics}

Qumodes can also be used to probe the thermodynamics of quantum systems that are perturbed far from equilibrium. In particular, we can probe the average work performed on a system due to a sudden quench in the interaction Hamiltonian -- and moreover, determine the irreversible portion of this work $\langle W_{\text{irr}}\rangle$~\cite{tasaki2000jarzynski, campisi2011colloquium, dorner2012emergent}, defined as the difference between $\langle W\rangle$, average work done on the system during the quench, and $\Delta F$, the change in the free energy had the system evolved adiabatically from the thermal state of the initial interaction Hamiltonian to that of the final Hamiltonian. Interest in the irreversible component of work is motivated by the fluctuation theorems and its connections with various entropy measures~\cite{deffner2010generalized, tasaki2000jarzynski} -- it is a widely-used measure of irreversibility, and has been shown to be a signature for some second-order phase transitions~\cite{mascarenhas2014}. Note that the average work is also an interesting quantity to study in its own right, and its behaviour across a critical point has been connected to first-order quantum phase transitions~\cite{mascarenhas2014}.  

The means of measuring average work and its irreversible component proceeeds as follows. Consider initial and final interaction Hamiltonians $H_{\text{int}}^{(0)}$ and  $H_{\text{int}}^{(1)}$, the former proportional to the system Hamiltonian $H_\Theta$, and the latter satisfying $\lambda H_{\text{int}}^{(1)}\gg H_\Theta$. For each of the interaction Hamiltonians we can apply the methods above to determine the free energies of their respective associated thermal states $\rho_{\Theta_i}$, and correspondingly, the difference in these free energies $\Delta F$. Under a sudden quench, the system is unchanged and will remain in its initial thermal state $\rho_{\Theta}$, which is equivalent to a thermal state of $H_{\text{int}}^{(0)}$ at some effective inverse temperature $\tilde{\beta}$. The proportionality between the system Hamiltonian $H_\Theta$, and the initial interaction Hamiltonian $H_{\mathrm{int}}^{(0)}$ corresponds to the ratio between their eigenvalues $E_n/\bar{E}_n$, which also corresponds to the ratio of their associated temperatures $\tilde{T}/T=\beta/\tilde{\beta}$.

Thus, the average work done by the quench is given by 
\begin{align}
\langle W \rangle&=\text{Tr}(\rho_{\Theta}H_{\text{int}}^{(1)})-\text{Tr}(\rho_{\Theta}H_{\text{int}}^{(0)})\nonumber\\
&=\sum_{m} E^{(1)}_m P^{(1)}_m-\sum_n E^{(0)}_n P^{(0)}_n,
\end{align}
where $\{E^{(i)}_n\}$ and $\{P^{(i)}_n\}$ are energies and probability amplitudes associated with the state $\rho_{\Theta_0}$ for each of the interaction Hamiltonians $H_{\text{int}}^{(i)}$. The second term corresponds to the average energy of the state with respect to the initial Hamiltonian, and so may be determined in the initial stage when we are calculating the associated free energy. The first term can be measured by using the qumode probing protocol to determine the $\{E^{(1)}_n\}$ and $\{P^{(1)}_n\}$. Finally, we can calculate the irreversible component of the average work through $\langle W_{\text{irr}}\rangle=\langle W \rangle-\Delta F$.

Finally, we remark that we can also use qumode probes to find the overlaps of the ground states of a parameter-dependent Hamiltonian at two different values of the parameter $\gamma$. Such overlaps have been used to characterise regions of criticality defining quantum phase transitions, such as in the Dicke model~\cite{zanardi2006ground}. Let $H_{\text{int}}(\gamma)$ denote the Hamiltonian when the parameter takes on a particular value $\gamma$, and let $\gamma_c$ be the value of the parameter at the critical point. The overlap of the ground states $\ket{u_\gamma}$ associated with two values of the parameter $\gamma_0<\gamma_c$ and $\gamma_1>\gamma_c$ can be found using a concatenation of two qumode probing circuits. The first probing circuit is run using $H_{\mathrm{int}}(\gamma_0)$, and is post-selected on the qumode measurement resulting in the initial ground state energy, such that the state is left in the ground state of this Hamiltonian $\ket{u_{\gamma_0}}$. We then probe this state using a second qumode, coupled with the interaction Hamiltonian $H_{\mathrm{int}}(\gamma_1)$. The probability that the qumode is found to have been shifted according to the ground state of this latter Hamiltonian then corresponds to $|\braket{u_{\gamma_0}}{u_{\gamma_1}}|^2$, i.e., the desired overlap probability. 

\section{Example: Spin-1/2 Particle in Transverse Field}

Consider a two level system with energies $\pm\Delta$, for example, a spin-1/2 particle within a magnetic field of strength $\Delta$ along the $z$-direction. Consider further that the spin is subject to an additional field of strength $B$ along the $x$-axis, such that the total Hamiltonian is given by
\begin{equation}
H_{\mathrm{sys}}=\Delta\sigma_z+B\sigma_x.
\end{equation}

We now illustrate thermometry with a qumode by applying it to this example, when the system is thermalised to this Hamiltonian at a range of temperatures $T$. Specifically, we simulate sampling from the distribution obtained by employing the probing protocol with a qumode initially prepared in a squeezed momentum state. We study a scenario where the interaction Hamiltonian is proportional to the system Hamiltonian, i.e., $H_{\mathrm{int}}\propto H_{\mathrm{sys}}$. 

We will also, as an extra test of the protocol, mimic that the eigenvalues of $H_{\mathrm{sys}}$ and $H_{\mathrm{int}}$ are unknown (but their relative proportionality is), and demand that the protocol also deduce these. We assume that it is known that there is one positive and one negative energy eigenvalue, such that the protocol will treat all measurements corresponding to a positive shift in momentum as being due to the negative eigenvalue, and vice versa. These are then averaged over to arrive at the deduced energy eigenvalues.

\begin{figure*}
\includegraphics[width=0.95\linewidth]{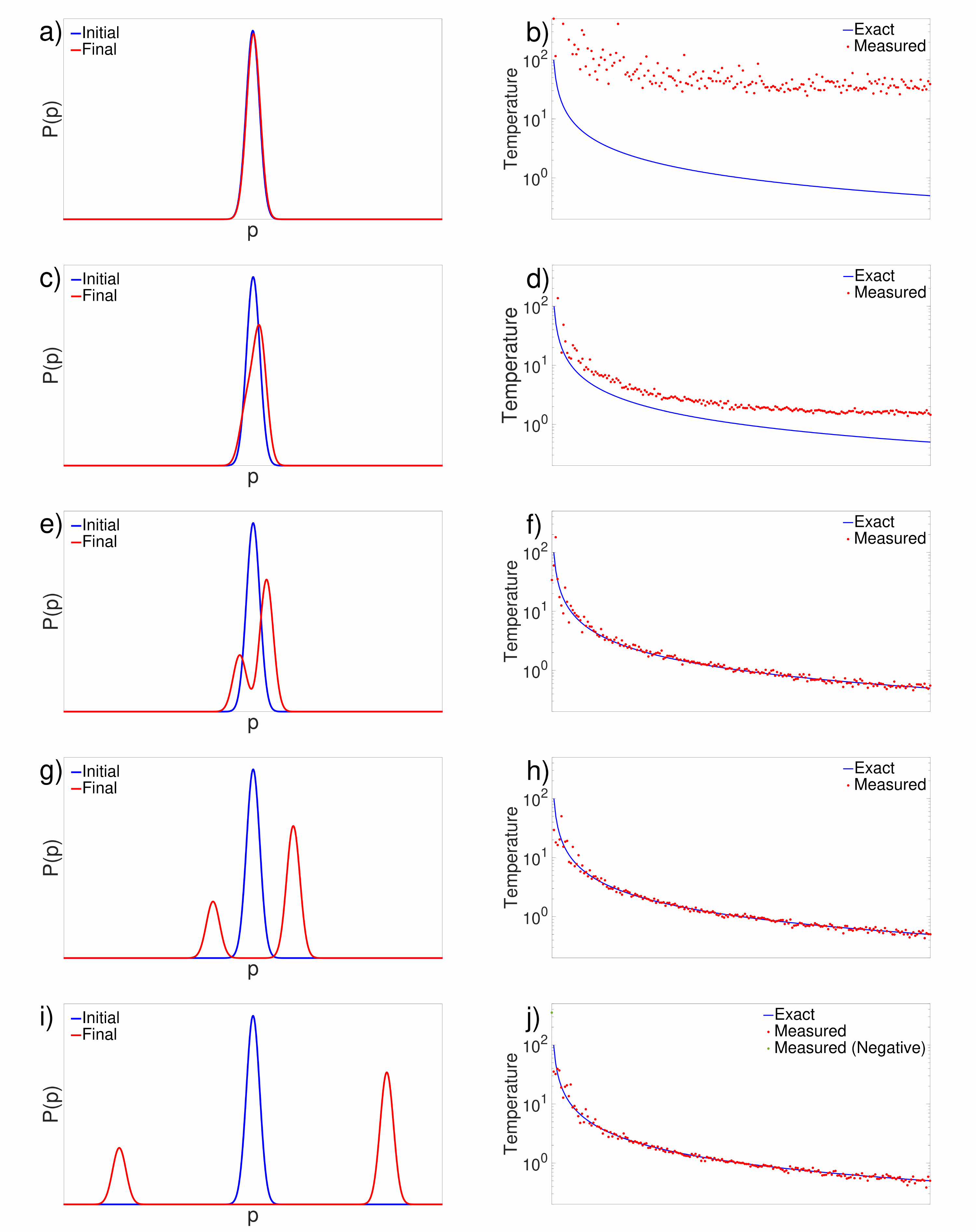}
\caption{{\bf Simulated thermometry of spin-1/2 in tranverse field.} (a,c,e,g,i) Initial and final momenta distributions of qumodes used to probe the temperature of a spin-1/2 particle (at $\beta=0.3$), and (b,d,f,h,j) qumode-measured temperature with $10^3$ shots for various $\beta\in[0,2]$. Precisions $\alpha\in\{0.1,0.5,1,3,10\}$ correspond to plots $\{(a,b),(c,d),(e,f),(g,h),(i,j)\}$ respectively. One temperature measurement yielded a negative value for $\alpha=10,\beta=0$; the corresponding absolute value is shown.}
\label{figspin}
\end{figure*}

We examine the case that $\Delta=B=1$, $H_{\mathrm{int}}=H_{\mathrm{sys}}$. We consider temperatures in the range $\beta=1/T\in[0,2]$, with the extremes corresponding to the infinite temperature limit ($P_0=P_1$), and a very low temperature limit ($P_0\approx 0.997$, $P_1\approx0.003$). We allow the qumode to make $10^3$ measurements (`shots') of the distribution, and combine the dependence on squeezing $s$, coupling strength $\lambda$ and interaction time $\tau$ in a single precision factor $\alpha:=s\lambda\tau$. To illustrate the effect of the precision, we plot the final distributions of the qumode state for $\alpha\in\{0.1,0.5,1,3,10\}$ at $\beta=0.3$ in \figsref{figspin}(a,c,e,g,i) respectively.

The performance of the qumode at estimating the temperature of the spin for the prescibed range of temperatures is shown in \figsref{figspin}(b,d,f,h,j) for each of the respective precisions. We see that at low precision the method overestimates the temperature. This is because the shift of the qumode momentum during the interaction is much smaller than the finite width of the initial distribution, and so a significant portion of the tails of the distribution are erroneously assigned to the wrong eigenvalue, creating the illusion of a more balanced population of the two states. We also observe that generally the accuracy is lower at high temperature (low $\beta$); this is because the exponential dependence of the state probabilities on $\beta$ renders a high degree of sensitivity in the deduced $\beta$ with respect to the measured probabilities at high temperature (where $P_0\approx P_1$). Indeed, when the measured probability for the excited state is greater than 1/2 (even if only to an arbitrarily small degree), the deduced temperature is negative. Ultimately, this can be ameliorated by using a larger number of shots that reduce the magnitude of the fluctuations in the measured probabilities. We remark that further numerical study (not shown) indicates that the number of shots is the limiting factor in the accuracy (rather than precision) for the shown plots at larger precision, and a visibly tighter clustering around the exact temperatues is observed for the measured temperatures at $\alpha=10$ when the number of shots is increased to $10^4$.

\section{Conclusion}

Properties of quantum systems may be imprinted onto continuous variable qumode ancillae, allowing for non-destructive probing of the system. For an appropriate choice of interaction operator and initial qumode state, the occupation probabilities of the desired observable's eigenstates with respect to the system state are mapped directly on to the qumode state. The qumode state then behaves according to the same statistics as this operator, and thus measurement of the qumode reproduces the same result as a direct measurement of the system would, while avoiding particular drawbacks associated with direct measurements of practical implementations of quantum technologies. We have shown that this method of probing has a strong potential for use in probing the thermodynamics of quantum systems, allowing access to many quantities of interest, particularly in the context of cold atom quantum simulators.

For systems in equilibrium, qumode probing can be used for thermometry of the system temperature, as well as reconstruction of the partition function, from which many other thermodynamical quantities of interest can be deduced. Moreover, for systems outside of equilibrium, qumodes can be used to probe work and free energy differences under evolutions of the system and its Hamiltonian. As the basic form of the interaction is almost synonymous with the form of quantum light-matter interactions, there is a significant scope of applicability of these results, particularly given that the necessary parameter regimes can be achieved in current experiments. Indeed, we remark that while several of our results require the system component of the interaction Hamiltonian to be proportional to the thermalising Hamiltonian of the system, this is not an unrealistic challenge particularly in systems driven by their interactions with light or other similar environments, especially in the strong-coupling regime. With the high degree of tunability of the specific form of the interaction offered by optical setups, qumode probing promises to be a valuable tool in the experimental characterisation of many-body quantum systems.

\acknowledgments The authors acknowledge financial support from Grants No. FQXi-RFP-1809 and No. FQXi-RFP-IPW-1903 from the Foundational Questions Institute and Fetzer Franklin Fund (a donor advised fund of Silicon Valley Community Foundation), the University of Manchester Dame Kathleen Ollerenshaw Fellowship, the Imperial College Borland Fellowship in Mathematics, the Singapore Ministry of Education Tier 2 Grant MOE-T2EP50221-0005, the Singapore Ministry of Education Tier 1 Grant RG77/22, and the National Research Foundation, Singapore, and Agency for Science, Technology and Research (A*STAR) under its QEP2.0 programme (NRF2021-QEP2-02-P06, and funding from the Science and Technology Program of Shanghai, China (21JC1402900). Any opinions, findings and conclusions or recommendations expressed in this material are those of the authors and do not reflect the views of National Research Foundation or the Ministry of Education Singapore. T.~J.~E.~and N.~L.~thank the Centre for Quantum Technologies for their hospitality.

{\bf Data Availability.} Data sharing not applicable to this article as no datasets were generated or analysed during the current study.

{\bf Author Contributions.} All authors contributed to the research and writing of the manuscript.

{\bf Competing Interests.} The authors declare no competing interests.

\bibliography{ref}

\end{document}